# A HIDDEN SPATIAL-TEMPORAL MARKOV RANDOM FIELD MODEL FOR NETWORK-BASED ANALYSIS OF TIME COURSE GENE EXPRESSION DATA[1]

By Zhi Wei and Hongzhe Li

*University of Pennsylvania School of Medicine*

Microarray time course (MTC) gene expression data are commonly collected to study the dynamic nature of biological processes. One important problem is to identify genes that show different expression profiles over time and pathways that are perturbed during a given biological process. While methods are available to identify the genes with differential expression levels over time, there is a lack of methods that can incorporate the pathway information in identifying the pathways being modified/activated during a biological process. In this paper we develop a hidden spatial-temporal Markov random field (hstMRF)-based method for identifying genes and subnetworks that are related to biological processes, where the dependency of the differential expression patterns of genes on the networks are modeled over time and over the network of pathways. Simulation studies indicated that the method is quite effective in identifying genes and modified subnetworks and has higher sensitivity than the commonly used procedures that do not use the pathway structure or time dependency information, with similar false discovery rates. Application to a microarray gene expression study of systemic inflammation in humans identified a core set of genes on the KEGG pathways that show clear differential expression patterns over time. In addition, the method confirmed that the TOLL-like signaling pathway plays an important role in immune response to endotoxins.

**1. Introduction.** Cellular activities are often dynamic and it is therefore critical to study the gene expression patterns over time in biology. With the advances in high throughput gene expression profiling technologies, microarray time course (MTC) experiments remain a common tool to cap-

Received May 2007; revised October 2007.
[1]Supported by NIH Grant CA127334 and a grant from the Pennsylvania Department of Health.

ture the gene expression patterns over time in a genomic scale. This is evidenced by the fact that such MTC data account for more than one third of gene expression studies in the Gene Expression Omnibus, a database repository of high throughput gene expression data hosted by the National Center for Biotechnology Information (http://www.ncbi.nlm.nih.gov/geo/). One important application of such MTC gene expression experiments is to identify genes that are temporally differentially expressed (TDE) between two MTC experiments and the pathways or networks that are perturbed or activated during a given biological process. Compared to gene expression studies at one time point, such MTC studies can potentially identify more genes that are differentially expressed [Yuan and Kendziorski (2006), Tai and Speed (2006) and Hong and Li (2006)].

One important feature of the MTC gene expression data is that the data are expected to be dependent over time. Efficiently utilizing such dependency can lead to a gain in efficiency in identifying the TDE genes. Several new statistical methods have been developed for identifying the TDE genes to account for such dependency. Storey et al. (2005) developed a method using basis function expansion to characterize the time-course gene expression data and proposed to develop gene-specific summary statistics and the corresponding $p$ values based on the fitted smooth curves. Yuan and Kendziorski (2006) proposed to use a hidden Markov model to identify TDE genes in order to utilize the dependency of differential expressions of genes across time points. Tai and Speed (2006) developed the empirical Bayes method treating the observed time-course gene expression data as multivariate vectors. Hong and Li (2006) developed a functional empirical Bayes method using B-splines.

Although these new methods can be used to identify the TDE genes, they often do not provide direct information on which key molecular mechanisms are involved in the biological process or which biological pathways are being activated or modified during a given biological process. It is therefore important to develop novel statistical methods for identifying these TDE genes in the context of known biological pathways. Information about gene regulatory dependence has been accumulated from many years of biomedical experiments and is summarized in the form of pathways and assembled into pathway databases. Some well-known pathway databases include KEGG [Kanehisa and Goto (2002)], Reactome (www.reactome.org), BioCarta (www.biocarta.com) and BioCyc (www.biocyc.org). Several methods have recently been developed to incorporate the pathway structures into analysis of microarray gene expression data. Subramanian et al. (2005) developed a gene set enrichment analysis procedure to account for the group structure of genomic data and to identify pathways that are related to diseases or biological processes. Rahnenführer et al. (2004) demonstrated that the sensitivity of detecting relevant pathways can be improved by integrating information about pathway topology. In Sivachenko et al. (2005) a network topology extracted from the literature was used jointly with microarray data to find significantly affected pathway regulators. Nacu et al. (2006) proposed an interesting permutation-based test for identifying subnetworks from a known network of genes that are related to phenotypes. Rapaport et al. (2007) proposed to first smooth the gene expression data on the network based on the spectral graph theory and then to use the smoothed data for classification. However, none of these explicitly models the MTC expression data.

Wei and Li (2007) have recently developed a hidden Markov random field (hMRF) model for identifying the subnetworks that show differential expression patterns between two conditions, and have demonstrated that the procedure is more sensitive in identifying the differentially expressed genes than those procedures that do not utilize the pathway structure information. In this paper, to efficiently identify the TDE genes in the MTC experiments, we develop the hMRF model [Wei and Li (2007)] further into the hidden spatial-temporal MRF (hstMRF) to simultaneously consider the spatial and temporal dependencies of differential expression states of genes. The key of our approach is that the information of a known network of pathways is efficiently utilized in the analysis of MTC expression data in order to identify more biologically interpretable results. We also present an algorithm that combines the features of the iterative conditional modes (ICM) algorithm of Besag (1986) and the Viterbi algorithm [Rabiner (1989)].

We introduce the hstMRF model in Section 2 and present an efficient algorithm for parameter estimation by the ICM algorithm and the Viterbi algorithm in Section 3. We present results from simulation studies in Section 4 to demonstrate the application of the hstMRF model, compare with existing methods and to evaluate the sensitivity of the method to misspecification of the network structure. In Section 5, for a case study, we apply the hstMRF model to analyze the MTC data of a systemic inflammation study in humans [Calvano et al. (2005)]. We present a brief discussion in Section 6.

**2. Statistical models and methods.** Consider the MTC gene expression data measured under two different conditions over time points $0, 1, 2, \ldots, T$. Let $\mathbf{y}_t$ be a $p \times (m+n)$ matrix of expression values for $p$ genes with $m+n$ arrays at time $t$, where the first $m$ columns are the expression data measured under the first condition over $m$ independent samples and the second $n$ columns are the expression data measured under the second condition over $n$ independent samples. The full set of observed expression values is then denoted by

$$\mathbf{Y} = (\mathbf{y}_0, \mathbf{y}_1, \ldots, \mathbf{y}_T).$$

With slight abuse of notation, let $\mathbf{y}_g$ denote one row of this matrix containing data for gene $g$ over time, $\mathbf{y}_{gt}$ denote expression data for gene $g$ at time $t$ over $m+n$ samples, and $y_{gtc}$ denote the expression level for gene $g$ at time $t$ in sample $c$. Suppose that we have a network of known pathways which can be represented as an undirected graph $G = (V, E)$, where $V$ is the set of nodes that represent genes or proteins coded by genes and $E$ is the set of edges linking two genes with a regulatory relationship. As an example, Figure 1 shows the Toll-like receptor signaling pathway in the KEGG database, where the squares are the genes (or, more precisely, the gene products) or gene clusters and the directed lines between two genes indicate some regulatory relationships between them. In this paper we do not consider the direction of the edges and only treat this pathway as an undirected graph. Some components of this pathway are also components of other pathways such as the MAPK and JAK-STAT signaling pathways. In the current KEGG database, there are a total of 33 such regulatory pathways that form a regulatory network.

Let $p = |V|$ be the number of genes that this network contains. Note the gene set $V$ is often a subset of all the genes that are probed on the gene expression arrays. If we want to include all the genes that are probed on the expression arrays, we can expand the network graph $G$ to include isolated nodes, which are those genes that are probed on the arrays but are not part of the known biological network. For two genes $g$ and $g'$, if there is a known regulatory relationship, we write $g \sim g'$. For a given gene $g$, let $N_g = \{g' : g \sim g' \in E\}$ be the set of genes that have a regulatory relationship with gene $g$. Our goal is to identify the genes on the network $G$ that are differentially expressed at each time point during the time course of a given biological experiment. Let $X_{gt}$ be the random variable that assigns a differential expression state (DES) to gene $g$ at time $t$, taking a value of 1 if the $g$th gene is differentially expressed (DE) at time $t$ or a value of 0 if it is equally expressed (EE) at time $t$. Let $\mathbf{X} = (\mathbf{x}_0, \ldots, \mathbf{x}_T)$, where $\mathbf{x}_t = (X_{1t}, \ldots, X_{pt})'$. We denote $\mathbf{X}^*$ as the true but unknown differential expression state and interpret this as a particular realization of the random matrix $\mathbf{X}$. Our goal is to recover the true but unobservable $\mathbf{X}^*$ from the observed data $\mathbf{Y}$. Using the Bayes formula, we propose to estimate $\mathbf{X}^*$ that maximize $Pr(\mathbf{X}|\mathbf{Y}) \propto Pr(\mathbf{Y}|\mathbf{X})Pr(\mathbf{X})$, the posterior density for the gene expression states $\mathbf{X}$, given the gene expression levels $\mathbf{Y}$ where $Pr(\mathbf{Y}|\mathbf{X})$ represents the evidence from the microarray experiments and the prior $Pr(\mathbf{X})$ represents our prior knowledge on gene regulation as provided by the gene network $G$. In the following, we first specify probability models for $Pr(\mathbf{X})$ and $Pr(\mathbf{Y}|\mathbf{X})$.

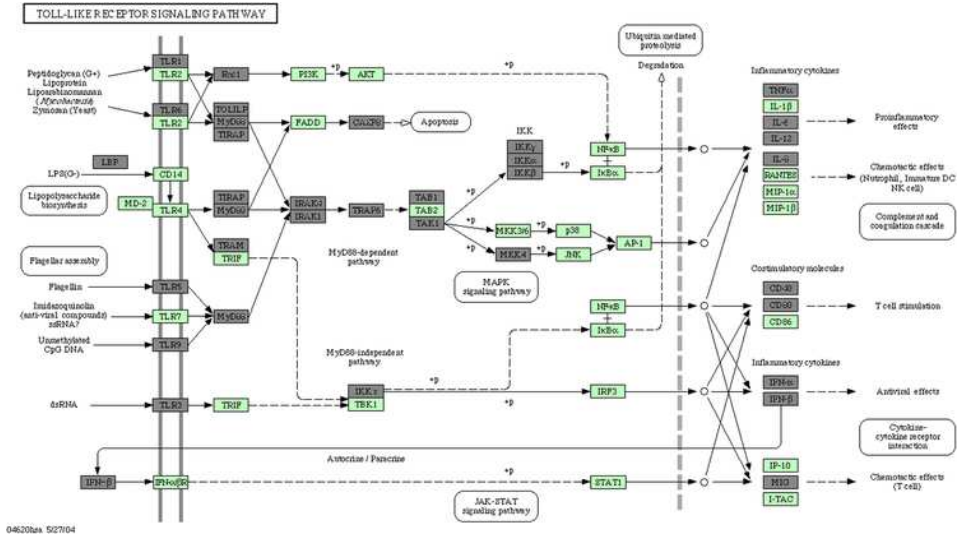

FIG. 1. *Structure of the KEGG Toll-like receptor (TLR) signaling pathway and the gene expression states of the genes at 4 h after the administration of endotoxin, where the EE genes are dark-shaded and the DE genes are light-shaded. In this pathway, the rectangles represent gene products, mostly proteins, but also including RNAs and the rounded rectangles indicate other pathways that are related to the TLR pathway. For the edges, a solid arrow between two proteins means activation, while a dashed arrow means indirect effect, and the "+p" sign on the arrow further specifies that activations are achieved by Phosphorylation. In addition, a rectangle pointing to a small circle, which points to another rectangle, means gene expression relationship. For example, the transcription factor STAT1 controls the mRNA expression of MIG. Finally, the two vertical lines on the left represent the cell membranes that separate cytoplasm and extracellular components and the vertical dashed line on the right represents the cell nuclear membrane. More detailed explanations of the KEGG pathways can be found at http:// www. genome. jp/ kegg/ document/ help_ pathway. html .*

2.1. *A spatial-temporal MRF model for the prior $Pr(\mathbf{X})$ on the network.* In order to define our proposed hstMRF model, we first specify the probability model for the latent differential expression states $\mathbf{X}$, taking into account both temporal dependency over time and the spatial dependency over the network. Specifically, for the initial time point 0, we define an auto-logistic model [Besag (1972, 1974)] as

$$(2.1) \qquad Pr(X_{g0}|X_{g'0}, g' \in N_g) = \frac{\exp\{X_{g0}F_1(X_{g0})\}}{1+\exp\{F_1(X_{g0})\}},$$

where $F_1(X_{g0}) = \gamma_0 + \beta_0 \sum_{g' \in N_g}(2X_{g'0} - 1)$, and $\gamma_0 \in \mathbf{R}$ and $\beta_0 \geq 0$. This model, which is equivalent to that assumed in Wei and Li (2007), assumes that at the initial time point 0, the conditional differential expression state of a given gene $g$ depends only on the differential expression states of its neighboring genes $N_g$.

For the following time points, we model $\{\mathbf{x}_t : t = 1, 2, \ldots, T\}$ by a $p$-dimensional vector Markov chain with the following transition probability:

$$
\begin{aligned}
(2.2) \quad & Pr(\mathbf{x}_t | \mathbf{x}_{t-1}) \\
& = \frac{1}{c_t} \exp \bigg\{ \gamma \sum_{g=1}^{p} X_{gt} + \beta_1 \sum_{g \sim g' \in E} (X_{gt} \oplus X_{g't}) \\
& \qquad\qquad + \beta_2 \sum_{g=1}^{p} (X_{gt} \oplus X_{g(t-1)}) \bigg\},
\end{aligned}
$$

where $c_t$ is the normalizing constant, $\oplus$ is the XNOR operator in the logic circuit, namely, it outputs 1 if the two inputs are the same, and 0 otherwise, $\gamma \in \mathbf{R}$ and $\beta_1 \geq 0$ and $\beta_2 \geq 0$ are the parameters that induce spatial and temporal dependencies of the differential expression states. From this transition probability, we can derive (see the Appendix for details of the derivation), for each gene $g$, the conditional distribution of $X_{gt}$ as

$$(2.3) \quad Pr(X_{gt} | \mathbf{x}_0, \mathbf{x}_1, \ldots, \mathbf{x}_{t-1}, X_{g't}, g' \neq g) = \frac{\exp\{X_{gt} F_2(X_{gt})\}}{1 + \exp\{F_2(X_{gt})\}},$$

where

$$F_2(X_{gt}) = \gamma + \beta_1 \sum_{g' \in N_g} (2X_{g't} - 1) + \beta_2 (2X_{g(t-1)} - 1).$$

Note that this conditional probability depends on the DES of its neighboring genes and the DES of this gene at the previous data point. Together, the initial distribution (2.1) and the conditional probability (2.3) define the probability distribution with parameter $\boldsymbol{\Phi} = (\gamma_0, \beta_0, \gamma, \beta_1, \beta_2)$ for the latent differential expression states.

2.2. *Gamma–Gamma model for observed gene expression data* $\mathbf{Y}$. To finish the specification of the hstMRF model, we also need to define the density function of the observed data $\mathbf{Y}$ given the latent DESs $\mathbf{X}$, $h(\mathbf{Y}|\mathbf{X})$. We make the following conditional independence assumption:

$$(2.4) \quad h(\mathbf{Y}|\mathbf{X}) = \prod_{t=0}^{T} \prod_{g=1}^{p} f(\mathbf{y}_{gt} | X_{gt}),$$

where $f(\mathbf{y}_{gt} | X_{gt})$ is the conditional density function of the observed expression values of $m + n$ samples for gene $g$ at time $t$ given the differential state $X_{gt}$. From the biological point of view, it is plausible to think that the differential expression states are more likely to be dependent over time. We therefore assume that the gene expression levels over time are independent given the differential expression states, which induce dependency of the gene expression levels over time. The same assumption was also made in Yuan and Kendziorski (2006) in their HMM formulation.

In order to specify $f(\mathbf{y}_{gt}|X_{gt})$, we propose to use the Gamma–Gamma (GG) model for gene expression data [Newton et al. (2001) and Kendiziorski et al. (2003)]. The same probability model was also used in Wei and Li (2007). Under such a Gamma–Gamma model, we assume that the observation $y_{gtc}$ is a sample from a gamma distribution having shape parameter $\alpha > 0$ and a mean value $\mu_g$, thus, with scale parameter $\lambda_i = \alpha/\mu_i$. Following Newton et al. (2001) and Kendiziorski et al. (2003), fixing $\alpha$, we assume that the quantity $\lambda_i = \alpha/\mu_i$ has a gamma distribution with shape parameter $\alpha_0$ and scale parameter $v$. Let $\boldsymbol{\Theta} = (\alpha, \alpha_0, v)$ be the parameters used to specify these two distributions. Under this hierarchical model, Kendziorski et al. (2003) derived the following conditional density function for the gene expression data:

$$(2.5) \quad f(\mathbf{y}_{gt}|X_{gt}; \boldsymbol{\Theta}) = \begin{cases} \dfrac{K_1 K_2 (\prod_{j=1}^{m+n} y_{gtj})^{\alpha-1}}{(v + y_{gt.m})^{m\alpha+\alpha_0}(v + y_{gt.n})^{n\alpha+\alpha_0}}, & \text{if } X_{gt} = 1, \\ \dfrac{K(\prod_{j=1}^{m+n} y_{gtj})^{\alpha-1}}{(v + y_{gt.m} + y_{gt.n})^{(m+n)\alpha+\alpha_0}}, & \text{if } X_{gt} = 0, \end{cases}$$

where

$$y_{gt.m} = \sum_{j=1}^{m} y_{gtj}, \qquad y_{gt.n} = \sum_{j=m+1}^{m+n} y_{gtj}$$

and

$$K_1 = \frac{v^{\alpha_0} \Gamma(m\alpha + \alpha_0)}{\Gamma^m(\alpha)\Gamma(\alpha_0)}, \qquad K_2 = \frac{v^{\alpha_0}\Gamma(n\alpha + \alpha_0)}{\Gamma^n(\alpha)\Gamma(\alpha_0)},$$

$$K = \frac{v^{\alpha_0}\Gamma((m+n)\alpha + \alpha_0)}{\Gamma^{m+n}(\alpha)\Gamma(\alpha_0)}.$$

Together, models (2.1), (2.2) and (2.4) define a hstMRF model for MTC gene expression data with parameters $\boldsymbol{\Phi}$ and $\boldsymbol{\Theta}$. Similar to the HMM approach of Yuan and Kendziorski (2006), the model also assumes that the expression states of one particular gene over time follow a hidden Markov chain. While all genes in the HMM approach follow the same HMM model, each gene in the hstMRF model has its own specific HMM determined by its regulatory neighboring genes. The hstMRF model reduces to the HMM when all the genes in the networks are independent. It is also clear that the hMRF model in Wei and Li (2007) is a special case of the hstMRF model, when there is only one time point.

**3. Parameter estimation using ICM and Viterbi algorithms.** We propose the following algorithm based on the ICM algorithm of Besag (1986) and the Viterbi algorithm [Rabiner (1989)] to estimate the $\mathbf{\Phi}$ parameter in the hidden spatial-temporal MRF model and the $\mathbf{\Theta}$ parameter in the Gamma–Gamma model. The algorithm involves the following iterative steps:

1. Obtain an initial estimate $\hat{\mathbf{X}}$ of the true state $\mathbf{X}^*$, using simple two sample t-tests at each time point.

2. Estimate $\mathbf{\Phi}$ by the value $\hat{\mathbf{\Phi}}$ which maximizes the following pseudolikelihood likelihood [Besag (1974)] $l(\hat{\mathbf{X}}; \mathbf{\Phi})$ based on the current $\hat{\mathbf{X}}$:

$$l(\mathbf{X}; \mathbf{\Phi}) = \prod_{g=1}^{p} \frac{\exp\{X_{g0} F_1(X_{g0})\}}{1 + \exp\{F_1(X_{g0})\}} \times \prod_{t=1}^{T} \prod_{g=1}^{p} \frac{\exp\{X_{gt} F_2(X_{gt})\}}{1 + \exp\{F_2(X_{gt})\}}.$$

Maximizing this equation could be processed to obtain the estimate $\hat{\mathbf{\Phi}}$ by a standard logistic regression software routine such as *glm* in R. The rationale of using the pseudolikelihood for updating the parameter $\mathbf{\Phi}$ is that it is difficult to evaluate the full likelihood due to an unknown normalizing constant in the likelihood function.

3. Estimate $\mathbf{\Theta}$ by the value $\hat{\mathbf{\Theta}}$, which maximizes the conditional likelihood $h(\mathbf{Y}|\hat{\mathbf{X}}; \mathbf{\Theta})$ [equation (2.4)].

4. Update $\hat{\mathbf{X}}$ based on the current $\hat{\mathbf{\Phi}}$ and $\hat{\mathbf{\Theta}}$ using a combination of the ICM algorithm and the Viterbi algorithm [Rabiner (1989)]. Suppose that $\hat{\mathbf{X}}$ is the current estimate of the true $\mathbf{X}^*$; our goal is to update the current DES $\mathbf{X}_{g\cdot} \equiv (X_{g0}, X_{g1}, \ldots, X_{gT})$ of gene $g$ in light of all available information. Specifically, we update $\mathbf{X}_{g\cdot}$ by maximizing the conditional probability with respect to $\mathbf{X}_{g\cdot}$, given the observed data $\mathbf{Y}$ and the current DES of all other genes $\hat{\mathbf{X}}_{V \backslash g}$. This conditional probability can be written as

$$Pr(\mathbf{X}_{g\cdot}|\mathbf{Y}, \hat{\mathbf{X}}_{V \backslash g}) \propto f(\mathbf{Y}_g|\mathbf{X}_{g\cdot}) Pr(\mathbf{X}_{g\cdot}|\hat{\mathbf{X}}_{V \backslash g}).$$

For a given gene $g$, the most probable $\mathbf{X}_{g\cdot}$ can be obtained using the Viterbi algorithm with the transition probability defined as in equation (2.3). When applied to each gene in turn, this procedure defines a single cycle of the ICM algorithm.

5. Go to step 2 for a fixed number of cycles or until approximate convergence of $\hat{\mathbf{X}}$.

As noted by Besag (1974), since

$$Pr(\hat{\mathbf{X}}|\mathbf{Y}) = Pr(\hat{\mathbf{X}}_{g\cdot}|\mathbf{Y}, \hat{\mathbf{X}}_{V \backslash g}) Pr(\hat{\mathbf{X}}_{V \backslash g}|\mathbf{Y}),$$

$Pr(\hat{\mathbf{X}}|\mathbf{Y})$ never decreases at any stage and eventual convergence is assured. In our implementation, we stop the iterations when the maximum of the relative changes of the parameter estimates is smaller than a small value $\epsilon$, which is set to be 0.01 for simulations and 0.002 for the real data analysis. The converged $\hat{\mathbf{X}}$ are then taken to be the estimate of the true differential expression states. These estimates can then be mapped back to the network to identify the subnetworks at a given time, which are defined as those connected genes that show differential expressions between the two experimental conditions. We can then examine these DES over time to obtain a temporal view of the subnetworks with respect to the differential expression states.

## 4. Simulation study.

4.1. *Performance and comparisons with other methods.* We conducted simulation studies to evaluate the proposed procedure and to compare the results with other procedures, including the HMM approach [Yuan and Kendziorski (2006)], which only takes into account the time dependency of differential expression states, and the hMRF model [Wei and Li (2007)], which only takes into account the dependency of the differential expression states on the network. We simulated data based on the regulatory network of pathways provided by KEGG. Specifically, 33 human regulatory pathways were obtained from the KEGG database [December (2006)] including only the gene–gene regulatory relations and excluding compound–gene and compound–compound relations. The remaining gene–gene regulatory data were represented as an undirected graph where each node represents a gene and two nodes are connected by an edge if there is a regulatory relation between them. Loops (nodes connected to themselves) were eliminated. This resulted in a graph with 1668 nodes and 8011 edges.

Our first simulation follows that of Yuan and Kendziorski (2006) where only the dependency of the differential expression states over time was simulated. Specifically, the GG mixture model is specified at each time point with parameter $\mathbf{\Theta} = (10, 0.9, 0.5)$ and transition probabilities of differential states over time are defined as

$$Pr(X_{gt} = DE | X_{g(t-1)} = DE) = 0.7, \qquad Pr(X_{gt} = DE | X_{g(t-1)} = EE) = 0.1$$

for $t = 1, \ldots, 5$ while for the first time point

$$Pr(X_{g0} = DE) = 0.1.$$

There are 6 time points in total and 3 replicates for each condition at each time point. We simulated 100 such datasets and each set contains 1,668 genes on the KEGG regulatory network. For the hstMRF and the hMRF approaches, we used the KEGG pathway structures in our analysis. The average sensitivity, specificity and the observed false discovery rate (FDR) for the proposed hstMRF, HMM and hMRF over the 100 simulated datasets at each of the 6 time points are shown in the first three columns of Table 1, where the sensitivity is calculated as the average over the 100 replications of the fraction of DE genes correctly identified by the method; specificity is the average of the EE genes correctly identified; and the false discovery rate (FDR) is the average of the ratio of the number of false positives to the number of the genes identified as DE. We observed that the proposed hstMRF model performs almost identically to the HMM model in sensitivity, specificity and FDR. In addition, both the hstMRF and the HMM procedure performed much better than the hMRF model; the increase of sensitivity can be more than 10% depending on the time points. Further, the increase in sensitivity does not greatly increase the observed FDR; the difference among different methods is within 2%.

The second simulation is similar to that in Wei and Li (2007), where only the spatial dependency of the DES was simulated using the hMRF model. For each time point, we randomly chose 9 pathways, initialized the genes in these pathways to be DE and the rest of the genes to be EE, and then we performed sampling five times iteratively conditional on the current sample of gene states to achieve the final sample of gene states according to equation (2.1) with $\gamma_0 = -2$ and $\beta_0 = 2$. Again, the GG mixture model with parameter $\Theta = (10, 0.9, 0.5)$ is assumed with three replicates in each condition. The results from different procedures are presented in the second three columns of Table 1. We observed that the hstMRF model performs similarly to the hMRF model and both procedures outperform the HMM in sensitivity at all time points and the increase of sensitivity can be as large as 14% depending on the time points. The increase in sensitivity does not involve an increase in the FDR. The hstMRF procedure has either considerably lower FDRs (time points 0, 1, 2 and 5) than HMM or comparable FDRs (time points 3 and 4).

The last simulation aims to simulate the differential states with both spatial and temporal dependencies. In particular, for the 33 KEGG pathways, we randomly picked 8 pathways at the 1st time point (time point 0) to be the DE pathway in which all the genes were initially set as DE genes. For time points 1 to 5, the DE/EE pathways were simulated according to the following transition probabilities:

$$Pr(\text{pathway}_{it} = DE|\text{pathway}_{i(t-1)} = EE) = 0.1,$$
$$Pr(\text{pathway}_{it} = DE|\text{pathway}_{i(t-1)} = DE) = 0.7.$$

Then for each simulated dataset, for each time point, we first set all genes in the DE pathways to be DE and then performed sampling five times based on the current gene states, according to equation (2.1) with $\gamma_0 = -2$ and $\beta_0 = 2$. The results from different procedures are presented in the last three columns of Table 1. We observed that the hstMRF model resulted in higher sensitivity and similar specificity in identifying the DE genes over time, as compared to the HMM or the hMRF models. The FDR rates are comparable to the HMM procedure with a slightly higher FDR rate at time point 3, 0.06 versus 0.023 and 0.027 for HMM and hMRF methods, respectively.

TABLE 1
*Comparison of performance in terms of sensitivity (SEN), specificity (SPE) and false discovery rate (FDR) of three different procedures based on 100 replications for three different scenarios. Standard errors range from 0.018 to 0.07 with a median of 0.027 for the sensitivity, from 0.001 to 0.004 with a median of 0.003 for the specificity and from 0.005 to 0.023 with a median of 0.009 for the FDR [see Supplementary materials for details, Wei and Li (2008)]. hstMRF: proposed hidden spatial-temporal Markov random field model; HMM: hidden Markov model; hMRF: hidden Markov random filed model*

|   |   | Temporal dependency | | | Spatial dependency | | | Spatial-temporal dependency | | |
|---|---|---|---|---|---|---|---|---|---|---|
|   |   | hstMRF | HMM | hMRF | hstMRF | HMM | hMRF | hstMRF | HMM | hMRF |
|   | t0 | 0.66 | 0.67 | 0.62 | 0.85 | 0.71 | 0.85 | 0.90 | 0.72 | 0.88 |
| S | t1 | 0.71 | 0.71 | 0.63 | 0.76 | 0.71 | 0.73 | 0.79 | 0.79 | 0.67 |
| E | t2 | 0.74 | 0.74 | 0.63 | 0.73 | 0.70 | 0.76 | 0.82 | 0.81 | 0.80 |
| N | t3 | 0.75 | 0.75 | 0.64 | 0.75 | 0.71 | 0.71 | 0.82 | 0.77 | 0.77 |
|   | t4 | 0.75 | 0.75 | 0.65 | 0.80 | 0.69 | 0.79 | 0.82 | 0.80 | 0.69 |
|   | t5 | 0.71 | 0.71 | 0.66 | 0.78 | 0.72 | 0.83 | 0.79 | 0.76 | 0.69 |
|   | t0 | 1.00 | 1.00 | 1.00 | 1.00 | 0.99 | 1.00 | 0.99 | 0.99 | 1.00 |
| S | t1 | 1.00 | 1.00 | 1.00 | 1.00 | 0.99 | 1.00 | 0.97 | 0.96 | 1.00 |
| P | t2 | 0.99 | 0.99 | 1.00 | 0.99 | 0.99 | 0.99 | 0.99 | 0.98 | 1.00 |
| E | t3 | 0.99 | 0.99 | 1.00 | 0.99 | 0.99 | 0.99 | 0.97 | 0.99 | 0.99 |
|   | t4 | 0.99 | 0.99 | 1.00 | 0.99 | 0.99 | 1.00 | 0.99 | 0.99 | 1.00 |
|   | t5 | 0.99 | 0.99 | 1.00 | 0.99 | 0.99 | 0.99 | 1.00 | 0.99 | 1.00 |
|   | t0 | 0.021 | 0.025 | 0.016 | 0.007 | 0.026 | 0.009 | 0.016 | 0.027 | 0.013 |
| F | t1 | 0.027 | 0.028 | 0.020 | 0.012 | 0.024 | 0.012 | 0.089 | 0.127 | 0.018 |
| D | t2 | 0.031 | 0.031 | 0.018 | 0.015 | 0.025 | 0.019 | 0.014 | 0.031 | 0.009 |
| R | t3 | 0.030 | 0.030 | 0.017 | 0.028 | 0.026 | 0.016 | 0.060 | 0.023 | 0.027 |
|   | t4 | 0.030 | 0.028 | 0.017 | 0.026 | 0.028 | 0.020 | 0.036 | 0.039 | 0.014 |
|   | t5 | 0.026 | 0.025 | 0.017 | 0.011 | 0.025 | 0.019 | 0.020 | 0.028 | 0.014 |

4.2. *Sensitivity to misspecification of the network structure.* Due to the fact that our current knowledge of biological networks is not complete, in practice, it is possible that the network structures that we use for network-based analysis are misspecified. The misspecification can be due to either the true edges of the networks being missed or the wrong edges being included in the network, or both of these two scenarios. We performed simulation studies to evaluate how sensitive the results of the hstMRF approach are to these three types of misspecifications of the network structures. We used the same datasets of 100 replicates as in the previous section (last 3 columns of Table 1), but used different misspecified network structures when we fitted the hstMRF model.

For the first scenario, we randomly removed 801 (10%), 2403 (30%) and 4005 (50%) from the 8011 true edges from the true KEGG networks when we fit the hstMRF model, respectively. For the second scenario, we randomly added approximately 801, 2403 and 4005 new edges to the KEGG network, respectively. Finally, for the third scenario, we randomly selected 90%, 70% and 50% of the 8011 true edges and also randomly added approximately 801, 2403 and 4005 new edges to the network, respectively, so that the total number of edges remained approximately 8011. The results of the simulations over 100 replications are summarized in Table 2. As expected, since the true number of DE genes is small, the specificities of the hstMRF procedure remain very high and are the same as when the true network structure was used. We also observed that the FDR rates also remained almost the same as when the true structure was used (see Table 1). However, we observed some decreases in sensitivity in identifying the true DE genes, especially at time point $t_0$. This is expected, since results in Table 1 indicate that, for the data we simulated, the network structure provides the most information for the DES at time point $t_0$. For other time points, temporal dependency contributes most of the information. It is worth pointing out that even when the network structure is largely misspecified as in scenario 3, the results from the hstMRF model are still comparable to those obtained from the HMM approach where the network structure is not utilized (see column 8 of Table 1). These simulations seem to indicate that the results of the hstMRF model are not too sensitive to the misspecification of the network structure unless the structure is greatly misspecified.

**5. Application to systemic inflammation gene expression study in humans.** We present results from an analysis of the systemic inflammation time course gene expression data in human whole blood leukocytes reported in Calvano et al. (2005), including time course gene expression profiles on eight healthy male and female subjects between 18 and 40 years of age. Using Affymetrix chips, Calvano et al. (2005) profiled the gene expression levels in human leukocytes immediately before (0 h) and at 2, 4, 6, 9 and 24 h after the intravenous administration of bacterial endotoxin for four healthy human subjects ($m = 4$, one female and three males). Four additional subjects ($n = 4$, one female and three males) without endotoxin administration were also profiled under identical conditions and were used as the controls. The robust multi-array (RMA) procedure [Irizarry et al. (2003)] was used to obtain the gene expression measures. To perform network-based analysis of the data, we merged the gene expression data with the 33 KEGG regulatory pathways and identified 1533 genes on the Hu133A chip that can be found in the 1668-node KEGG network of 33 pathways. Instead of considering all the genes on the Hu133A chip, we only focus analysis on these 1533 genes and aim to identify which genes and which subnetworks of the KEGG network of 33 pathways are perturbed or activated during the response to endotoxin.

TABLE 2
*Comparison of performance in terms of sensitivity (SEN), specificity (SPE) and false discovery rate (FDR) of the hstMRF procedure based on 100 replications when the network structure is misspecified. Standard errors range from 0.018 to 0.031 with a median of 0.024 for the sensitivity, from 0.002 to 0.007 with a median of 0.004 for the specificity and from 0.006 to 0.026 with a median of 0.011 for the FDR [see Supplementary materials for details, Wei and Li (2008)]. DEL: randomly deleting 10%, 30% and 50% of the true edges of the network; ADD: randomly adding approximately 801 (10%), 2403 (30%) and 4005 (50%) new edges to the network; DEL + ADD: randomly choosing 90%, 70% and 50% of the true edges and randomly adding 10%, 30% and 50% new edges to the network*

|   |   | 10% | | | 30% | | | 50% | | |
|---|---|---|---|---|---|---|---|---|---|---|
|   |   | **DEL** | **ADD** | **DEL + ADD** | **DEL** | **ADD** | **DEL + ADD** | **DEL** | **ADD** | **DEL + ADD** |
|   | t0 | 0.89 | 0.88 | 0.88 | 0.88 | 0.86 | 0.82 | 0.86 | 0.84 | 0.78 |
| S | t1 | 0.79 | 0.79 | 0.79 | 0.79 | 0.78 | 0.78 | 0.79 | 0.78 | 0.78 |
| E | t2 | 0.82 | 0.82 | 0.82 | 0.82 | 0.81 | 0.81 | 0.81 | 0.81 | 0.80 |
| N | t3 | 0.82 | 0.82 | 0.82 | 0.82 | 0.82 | 0.81 | 0.81 | 0.82 | 0.79 |
|   | t4 | 0.82 | 0.82 | 0.82 | 0.82 | 0.82 | 0.82 | 0.82 | 0.82 | 0.81 |
|   | t5 | 0.79 | 0.79 | 0.78 | 0.79 | 0.78 | 0.77 | 0.78 | 0.77 | 0.75 |
|   | t0 | 0.99 | 0.99 | 0.99 | 0.99 | 0.99 | 1.00 | 0.99 | 0.99 | 0.99 |
| S | t1 | 0.97 | 0.97 | 0.97 | 0.97 | 0.98 | 0.97 | 0.97 | 0.98 | 0.97 |
| P | t2 | 0.99 | 0.99 | 0.99 | 0.99 | 0.99 | 0.99 | 0.99 | 0.99 | 0.99 |
| E | t3 | 0.97 | 0.97 | 0.97 | 0.97 | 0.97 | 0.97 | 0.97 | 0.97 | 0.98 |
|   | t4 | 0.99 | 0.99 | 0.99 | 0.99 | 0.99 | 0.99 | 0.99 | 0.99 | 0.99 |
|   | t5 | 1.00 | 1.00 | 1.00 | 1.00 | 1.00 | 1.00 | 1.00 | 1.00 | 1.00 |
|   | t0 | 0.016 | 0.015 | 0.015 | 0.017 | 0.013 | 0.013 | 0.017 | 0.012 | 0.015 |
| F | t1 | 0.090 | 0.088 | 0.088 | 0.093 | 0.082 | 0.084 | 0.094 | 0.079 | 0.090 |
| D | t2 | 0.014 | 0.014 | 0.014 | 0.014 | 0.014 | 0.016 | 0.015 | 0.015 | 0.018 |
| R | t3 | 0.060 | 0.060 | 0.059 | 0.059 | 0.058 | 0.054 | 0.056 | 0.057 | 0.043 |
|   | t4 | 0.037 | 0.036 | 0.036 | 0.037 | 0.036 | 0.037 | 0.038 | 0.036 | 0.040 |
|   | t5 | 0.020 | 0.020 | 0.020 | 0.021 | 0.020 | 0.020 | 0.022 | 0.020 | 0.022 |

5.1. *Results from the hstMRF model.* The hstMRF model identified 35, 260, 326, 292, 258 and 127 DE genes at time points 0 h, 2 h, 4 h, 6 h, 9 h and 24 h, respectively. The parameter estimates were $\gamma_0 = -3.76$, $\beta_0 = 0.00001$, $\gamma = -0.71$, $\beta_1 = 0.013$ and $\beta_2 = 2.14$, indicating stronger time dependency than network dependency of gene differential expression states. The odds ratio in favor of being a DE gene is $\exp(2 \times 0.013) = 1.02$ if one of its neighbor genes is DE versus EE and $\exp(2 \times 0.013 \times 10) = 1.30$ if 10 of its neighboring genes are all DE versus all EE, conditional on the rest of the graph and assuming that all other DES are the same. In contrast, the odds ratio in favor of being a DE gene is $\exp(2 \times 2.14) = 72.24$ if this gene is a DE gene versus a EE at the previous time point assuming that the DES of its neighboring genes remain the same. A total of 362 unique DE genes were differentially expressed at least once at one of the six time points. Among these 362 DE genes, 262 of them are linked to at least one other gene on the KEGG network and 100 are isolated. The 326 DE genes at time point 4 h are from 31 out of the 33 pathways, indicating that the response to endotoxin administration in blood leukocytes can be viewed as an integrated cell-wide response. DAVID's enrichment analysis [Dennis et al. (2003)] showed that the three most significantly enriched pathways at time 4 h are the Toll-like receptor (TLR) signaling pathway, the Apoptosis pathway and the T cell receptor signaling pathway with the $p$-values of $4.2 \times 10^{-5}$, $8.1 \times 10^{-4}$ and $2.1 \times 10^{-3}$, respectively. As a comparison, at the 2 h time point, TLR pathway was ranked only 6th with a $p$-value 0.065. Such an increase in the TLR signaling pathway's significance is consistent with its well-known critical role in innate immunity [Aderem and Ulevitch (2000), Takeda et al. (2003) and Han and Ulevitch (2005)].

To demonstrate the spatial dependency of the DES of genes on the TLR pathway, Figure 1 presents the structure of the KEGG TLR pathway and the DES of the genes on this pathway at 4 h after the endotoxin administration, in which the DE genes are labeled in light shade. On this pathway, invading bacterial factors such as lipopolysaccharides (LPS, endotoxin) activate innate immunity, as well as stimulate the antigen-specific immune response and trigger the inflammatory response [Takeda et al. (2003)]. The signals stimulated by these factors are recognized by CD14, which in turn activates TLR4. MD-2 is a secreted protein that binds to the extracellular domain of TLR4 and is important in its signaling [Takeda et al. (2003) and Barton and Medzhitov (2003)]. Our analysis indicated that these three genes are differentially expressed together with TLR2 receptor. It is now understood that all TLRs activate a common signaling pathway that culminates in the activation of nuclear factor $\kappa$B (NF-$\kappa$B), as well as the mitogen-activated protein kinases (MAPKs) extracellular signal-regulated kinase (ERK), p38 and c-Jun N-terminal kinase (JNK) [Barton and Medzhitov (2003)]. This fact is clearly demonstrated by our analysis results: we observed that MKK36, NF-$\kappa$B, p38 and JNK were all differentially expressed. After the signals arrive at the transcriptional factors AP-1 and NF-$\kappa$B, they are activated and translocated into the nucleus. We observed that the down-stream genes of these transcription factors, including inflammatory cytokines (IL-1$\beta$, RANTES, MIP-1$\alpha$, MIP-1$\beta$), costimulatory molecules (CD86) were differentially expressed, consistent with the activation of innate immunity after administration of endotoxin [Aderem and Ulevitch (2000), Takeda et al. (2003) and Han and Ulevitch (2005)].

However, we did not observe differential expression of genes of the IL-1 receptor-associated kinase (IRAK) family, including the serine–threonine kinases IRAK1 and IRAK4 which are involved in the phosphorylation and activation of tumor necrosis factor (TNF) receptor-associated factor 6 (TRAF6), which was also not differentially expressed. Analysis of cells from mice lacking MyD88 has demonstrated that TLR4 is capable of inducing certain signaling pathways independent of the MyD88 adaptor [Takeda et al. (2003) and Barton and Medzhitov (2003)]. It is interesting to note that our analysis indicated that the Toll/interleukin-1 receptor (TIR) domain-containing adaptor-inducing IFN-$\beta$ (TRIF), which functions downstream of TLR4 [Barton and Medzhitov (2003)], was differentially expressed. TRIF is known to be responsible for the induction of interferon (IFN)-$\alpha$ and IFN-$\beta$ genes [Barton and Medzhitov (2003)], both of which were observed to be differentially expressed. The induction of IFN$\alpha/\beta$ genes by TLR4 further leads to activation of a key transcription factor interferon regulatory factor 3 (IFR3), which in turn led to differential expressions of the chemokines (IP-10, I-TAC). This suggests that the TRIF pathway may play an important role in response to endotoxin.

To further explore the temporal changes in KEGG subnetworks, we focused our analysis on the 262 connecting DE genes. We divided these 262 genes into nonoverlapping groups based on the first time point at which the gene became DE between the two groups, that is, the genes in group 1 are DE on 0 h, those in group 2 were DE on 2 h but not on 0 h, and those in group 3 were DE on 4 h but not on 0 h or 2 h. Other groups can be similarly defined. The genes that were DE at 24 h were also DE at least once at the previous time points. In addition, 9 DE genes were to be consistently over-expressed or under-expressed in the treatment group across all the time points. The remaining 253 genes include 9, 160, 70, 10 and 4 genes that were observed to first become DE at time points 0 h, 2 h, 4 h , 6 h and 9 h, respectively. We mapped these genes back to the KEGG gene network and showed the temporal response of gene expression in Figure 2 on the KEGG subnetwork. A clear temporal network response can be observed by highlighting the transient and self-limiting nature of this response. A large number of genes are differentially expressed from time point 2 h and the time point 4 h had the most number of the DE genes, which represents a quick response of the human immune system to the intrusion of endotoxin [Aderem and Ulevitch (2000) and Calvano et al. (2005)].

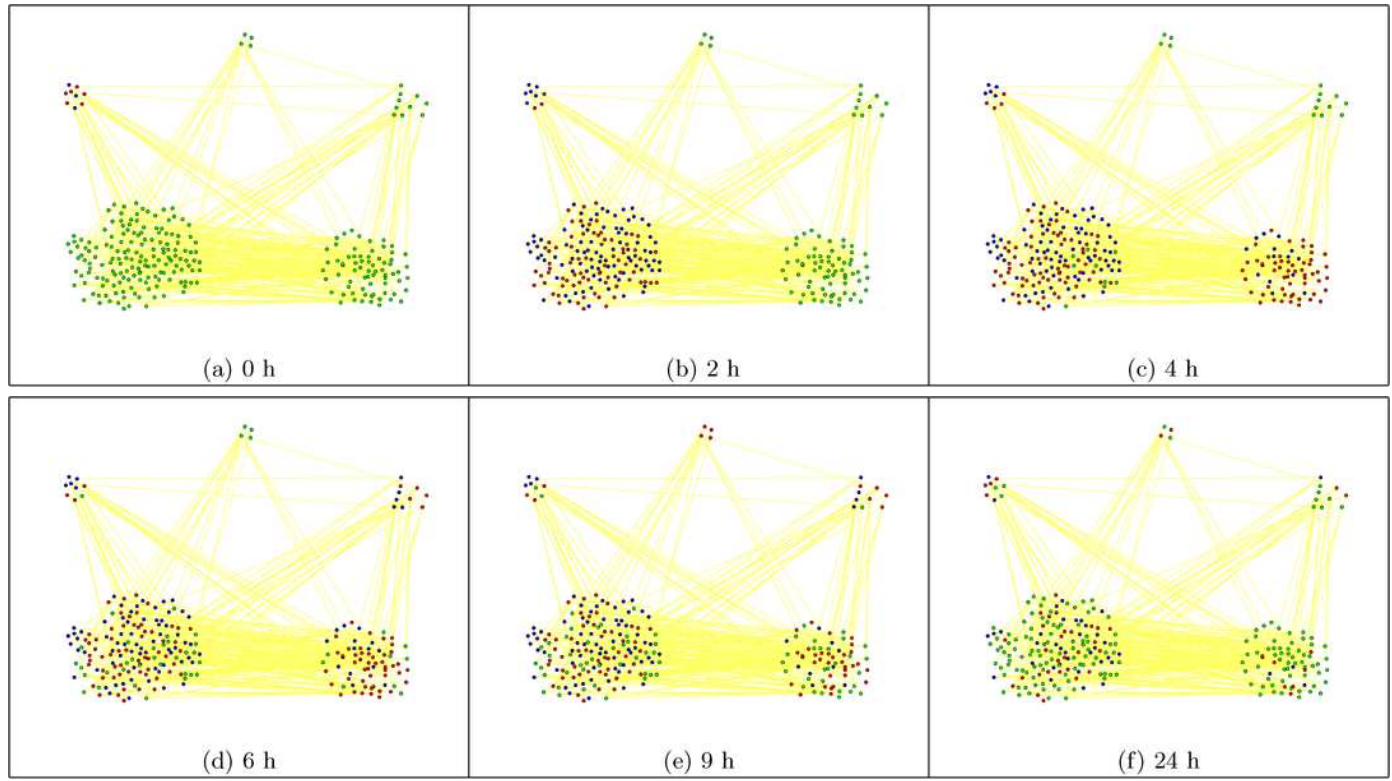

FIG. 2. *Temporal changes of gene expression levels in blood leukocytes on subnetworks of the KEGG pathways, showing a propagating and resolving procedure over time [0 h, 2 h, 4 h, 6 h, 9 h and 24 h from (a) to (f)]. Up-regulated genes in the endotoxin group are shown in red, down-regulated genes in the endotoxin group are shown in blue and equally expressed genes are shown in green.*

We also observed that a number of transcription factors were differentially expressed during time period 4–6 h after endotoxin injection, including both those that activate and those that inhibit the innate immune response. The activating genes included the signal transducer and activators of transcription genes (STAT1, STAT3, STAT4, STAT5A, STAT5B) and the inhibiting genes included the suppressor of cytokine signaling genes (SOCS1, SOCS2, SOCS3). There was also a delay (4–6 h) in increased mRNA abundance of secreted and membrane-associated proteins involved in the inflammatory response, including IL1RAP, IL1R2, IL1A, IL1B and IL1R1. Together, the temporal modulation of these DE genes controls the innate immune response in human leukocytes that progresses from an acute proinflammatory phase to unencumbered counter regulation, concluding with almost full recovery and a normal cellular state [Calvano et al. (2005)].

5.2. *Comparison with the results from the HMM.* As a comparison, the HMM assuming homogeneous transition probabilities identified 45, 227, 355, 342, 302 and 123 DE genes on the KEGG network at time points 0 h, 2 h, 4 h, 6 h, 9 h and 24 h, respectively. The DE genes identified at the 0 h and the 24 h are very similar between the two different approaches. Table 3 shows the number of DE and EE genes identified by the hstMRF and the HMM methods at 2 h, 4 h, 6 h and 9 h. While the sets of DE genes identified by the hstMRF and the HMM methods largely overlap, which is what we should expect because of the strong temporal effect, there are some differences in DE/EE genes identified, indicating the KEGG network structure indeed has impact on identifying the DE genes. At the 2 h after endotoxin administration, the hstMRF model identified 35 DE genes that were missed by the HMM. Plots of the average expression levels of these 35 genes at 0 h, 2 h and 4 h, indicating that most of them are differentially expressed at the 2 h [see Supplementary materials, Wei and Li (2008)]. As an example, Figure 3 shows the average expression levels of four of these genes at 0 h, 2 h and 4 h. One reason that the HMM did not identify these genes is that all these genes were at the EE state at time 0 h and the estimated transition probability from the EE state to DE state is only 0.06. In contrast, at the time points 4 h, 6 h and 9 h, there were 32, 49 and 43 DE genes identified by the HMM but missed by the hstMRF model, respectively. However, we observed that the HMM posterior probabilities of being a DE gene for these genes are relatively small, with a median value of 0.64, 0.64 and 0.67, respectively. In addition, we also observed that more than 75% of the neighboring genes of these DE genes are EE. The hstMRF model took into account the differential expression states of the neighboring genes in estimating the posterior probabilities and inferred these DE genes as the EE genes.

TABLE 3
*A comparison of the numbers of DE and EE genes identified by hstMRF and HMM at the 2 h, 4 h, 6 h and 9 h of the systemic inflammation gene expression experiments*

|     |    | hstMRF |     |      |     |      |     |      |     |
|-----|----|--------|-----|------|-----|------|-----|------|-----|
|     |    | 2 h    |     | 4 h  |     | 6 h  |     | 9 h  |     |
|     |    | EE     | DE  | EE   | DE  | EE   | DE  | EE   | DE  |
| HMM | EE | 1271   | 35  | 1173 | 5   | 1191 | 0   | 1231 | 0   |
|     | DE | 2      | 225 | 32   | 323 | 49   | 293 | 43   | 259 |

To further demonstrate the differences in genes identified by the HMM and the hstMRF methods, we performed analysis for data measured at the 0 h, 2 h and 24 h. The parameter estimates for the hstMRF model were $\beta_1 = 0.037$ and $\beta_2 = 0.37$, indicating less stronger temporal effects than our previous analysis. The hstMRF model identified 57 more DE genes at the 2 h than the HMM, of which 56 were EE at the 0 h and 24 h. Plots of the average expression profiles indeed show that all these 56 genes seem to show differential expression patterns at the 2 h and equally expression patterns at the 0 h and 24 h [see Supplementary materials, Wei and Li (2008)]. Under the hstMRF model, the DE neighboring genes increased the posterior probability of being a DE gene for these 56 genes. On average, these 56 DE genes have 2.4 more DE neighboring genes than what the EE genes have. Finally, it is interesting to note that 32 out of these 57 genes were identified as DE by the HMM if data from all the time points were used.

**6. Conclusion and discussion.** We have proposed a hidden spatial-temporal MRF model that utilizes the gene regulatory networks and temporal information simultaneously to identify DE genes in the analysis of microarray time course gene expression data. Simulation studies show that our methods outperform those methods capturing only regulatory dependence or capturing only time dependence in sensitivity, specificity and false discovery rate. We applied our method to analyze the MTC data of systemic inflammation in humans. The subpathways/subnetworks we identified at different time points show that the innate immune response in a human model progresses from an acute proinflammatory phase to unencumbered counter regulation, concluding with almost full recovery and a normal cellular state, consistent with the known characteristics of the human innate immune response [Aderem and Ulevitch (2000), Takeda et al. (2003) and Calvano et al. (2005)]. Our analysis also confirmed the critical role of the Toll-like receptor pathway in innate immune response and suggested that the signaling pathway during the human response to endotoxin might be through the TRIF pathway [Barton and Medzhitov (2003)].

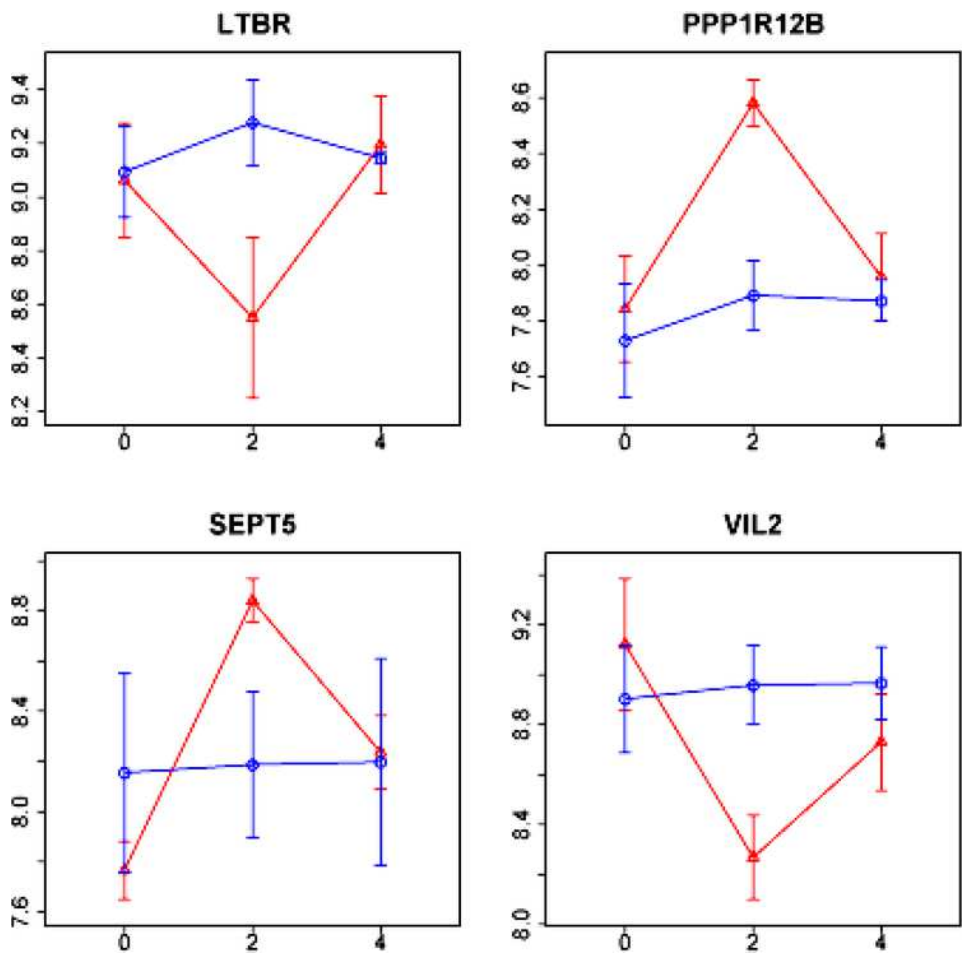

FIG. 3. *Average gene expression measures from the RMA procedure (in $\log_2$ scale) ($\pm 1$ SE) (y-axis) at the 0* h*, 2* h *and 4* h *(x-axis) for the four DE genes that were identified by the hstMRF model but missed by the HMM model.* $\Delta$*: group receiving the endotoxin administration; o: control group.*

In this paper we analyzed the systemic inflammation MTC data using KEGG pathways and aimed to identify the KEGG pathways affected by administration of endotoxin. However, the proposed methods can be applied to any other network of pathways. An important question is to decide which pathways one should use in analyzing the MTC data. This partially depends on the scientific questions to be addressed. If an investigator is only interested in a particular pathway, the proposed method can be applied to that particular pathway. If an investigator is interested in fully exploring his/her data and all available pathways, one should use a large collection of pathways, for example, the pathways collected by Pathway Commons (http://www.pathwaycommons.org/pc/) or build the network of pathways using some existing network construction tools [Basso et al. (2005)]. It should also be noted that our proposed methods can include all the genes probed on microarray by simply adding isolated nodes to the graphs. Another related issue is that our knowledge of pathways is not complete and can potentially include errors or misspecified edges on the networks. Although our simulations demonstrate that our methods are not too sensitive to the misspecification of the network structures, the effects of misspecification of the network on the results deserve further research. One possible solution to this problem is to first check the consistency of the pathway structure using the data available. For example, if the correlation in gene expression levels between two neighboring genes is very small, we may want to remove the edge from the pathway structure. Alternatively, one can build a set of new pathways using various data sources and compare these pathways with those in the pathway databases in order to identify the most plausible pathways for use in the proposed hstMRF method. Important future research will include how to represent and assess the uncertainty of the inference of the true differential expression states.

The proposed methods can be extended in several ways. First, besides the neighboring information, the pathways may provide additional biologically relevant information, such as inhibition and activation effects of genes and which genes are the transcriptional factors. The proposed methods treat all nodes and edges in the networks equally and use two parameters, $\beta_1$ and $\beta_2$, to characterize the spatial and temporal dependency of the differential states. One possible extension of the proposed methods is to incorporate the additional information about the pathways into data analysis. For example, we may attach more weight to transcription factors because of their immediate impact on mRNA production. Second, it is also possible to incorporate the promotor sequences and binding motif information of known transcription factors into the definition of the neighbors in our definition of the MRF models. Finally, since many networks are given by directed graphs, it is also possible to extend the MRF model to incorporate the direction of gene regulations.

In conclusion, microarray time course gene expression data are commonly collected to investigate the dynamic nature of important biological systems. The proposed methods facilitate the identification of the key molecular mechanisms involved and the cellular pathways being activated/modified during a given biological process. As our knowledge of the biological pathways increases, we expect more applications of such methods for identifying genes and pathways that are related to important biological processes.

## APPENDIX

We provide details on derivation of the conditional probability (2.3), which follows Zhu et al. (2005). First, from the definition of the transition probability (2.2), for any $1 \leq t \leq T$, we have

$$
\begin{aligned}
&Pr(\mathbf{x}_1,\ldots,\mathbf{x}_t|\mathbf{x}_0) \\
&\qquad = \prod_{t'=1}^{t} \frac{1}{c_{t'}} \exp\Bigg\{\sum_{t'=1}^{t}\bigg[\gamma\sum_{g=1}^{p} X_{gt'} + \beta_1 \sum_{g\sim g'\in E}(X_{gt'}\oplus X_{g't'}) \\
&\qquad\qquad\qquad\qquad\qquad\qquad + \beta_2 \sum_{g=1}^{p}(X_{gt'}\oplus X_{g(t'-1)})\bigg]\Bigg\}.
\end{aligned}
\qquad (A.1)
$$

From this, we have

$$
\begin{aligned}
&Pr(X_{gt}|\mathbf{x}_0,\mathbf{x}_1,\ldots,\mathbf{x}_{t-1},X_{g't},g'\neq g) \\
&= \frac{Pr(\mathbf{x}_1,\ldots,\mathbf{x}_t|\mathbf{x}_0)}{Pr(\mathbf{x}_1,\ldots,\mathbf{x}_{t-1},X_{gt}=0,X_{g't}|\mathbf{x}_0) + Pr(\mathbf{x}_1,\ldots,\mathbf{x}_{t-1},X_{gt}=1,X_{g't}|\mathbf{x}_0)} \\
&= \frac{\exp\{A + B(X_{gt})\}}{\exp\{A + B(0)\} + \exp\{A + B(1)\}},
\end{aligned}
$$

where

$$
A = \sum_{t'=1}^{t}\bigg[\gamma\sum_{g'\neq g} X_{g't'} + \beta_1 \sum_{g'\sim g''\in E\setminus\{g\}}(X_{g't'}\oplus X_{g''t'}) + \beta_2 \sum_{g'\neq g}(X_{g't'}\oplus X_{g'(t'-1)})\bigg],
$$

which consists of the terms in the exponent of (A.1) that do not include $X_{gt}$, where $E\setminus\{g\}$ is the set of the edges that do not include those that linked to gene $g$, and

$$
B(X_{gt}) = \gamma X_{gt} + \beta_1 \sum_{g'\in N_g}(X_{gt}\oplus X_{g't}) + \beta_2(X_{gt}\oplus X_{g(t-1)}),
$$

which consists of the terms that include $X_{gt}$. From this definition of $B(X_{gt})$, we have

$$B(0) = \beta_1 \sum_{g' \in N_g} (1 - X_{g't}) + \beta_2(1 - X_{g(t-1)}),$$

$$B(1) = \gamma + \beta_1 \sum_{g' \in N_g} X_{g't} + \beta_2 X_{g(t-1)}.$$

It is then easy to see that

$$\begin{aligned}
&Pr(X_{gt}|\mathbf{x}_0, \mathbf{x}_1, \ldots, \mathbf{x}_{t-1}, X_{g't}, g' \neq g) \\
&= \frac{\exp\{A + B(X_{gt})\}}{\exp\{A + B(0)\} + \exp\{A + B(1)\}} \\
&= \frac{\exp\{B(X_{gt}) - B(0)\}}{1 + \exp\{B(1) - B(0)\}} \\
&= \frac{\exp\{X_{gt}(\gamma + \beta_1 \sum_{g' \in N_g}(2X_{g't} - 1) + \beta_2(2X_{g(t-1)} - 1))\}}{1 + \exp\{(\gamma + \beta_1 \sum_{g' \in N_g}(2X_{g't} - 1) + \beta_2(2X_{g(t-1)} - 1))\}} \\
&= \frac{\exp\{X_{gt} F_2(X_{gt})\}}{1 + \exp\{F_2(X_{gt})\}},
\end{aligned}$$

which is the equation (2.3).

**Acknowledgments.** We thank four reviewers and the editor Dr. Michael Newton for many helpful comments that improved the presentation of the paper and Mr. Edmund Weisberg, MS, at Penn CCEB for editorial assistance.

## SUPPLEMENTARY MATERIAL

**Details on simulations and comparison with the HMM model** (doi: 10.1214/07-AOAS145SUPP; .pdf). We present detailed simulation results in Tables 1S-S2, including the standard errors of the sensitivities, specificities and FDRs. We also present the time course expression profiles (Figures S1-S2) of the genes that were identified by our methods but missed by the HMM method and the genes that were identified by the HMM method but missed by our methods.

Genomics and Computational Biology Graduate Group
Department of Biostatistics and Epidemiology
University of Pennsylvania School of Medicine
Philadelphia, Pennsylvania 19104
USA
E-mail: zhiwei@mail.med.upenn.edu
        hongzhe@mail.med.upenn.edu